\documentclass{webofc}
\usepackage[varg]{txfonts}   
\usepackage{epsfig}
\usepackage{graphicx}
\usepackage{mathptmx}      
\usepackage{bm}

\newcommand{\MeV}{\,{\mathrm{MeV}}}
\newcommand{\GeV}{\,{\mathrm{GeV}}}
\newcommand{\Km}{\,{\mathrm{Km}}}

\newcommand{\ignore}[1]{}

\wocname{EPJ Web of Conferences}
\woctitle{QCD@Work 2014}
\begin{document}
\title{Nonextensive thermodynamics with finite chemical potentials and protoneutron stars~\thanks{Presented by E.~Meg\'{\i}as at the QCD@Work: International Workshop on QCD, 16-19 June 2014, Giovinazzo, Bari, Italy.}
\fnsep
\thanks{Supported by Plan Nacional de Altas Energ\'{\i}as (FPA2011-25948), Junta de Andaluc\'{\i}a (FQM-225), Spanish Consolider-Ingenio 2010 Programme CPAN (CSD2007-00042), Spanish MINECO's Centro de Excelencia Severo Ochoa Program (SEV-2012-0234),  CNPq (Brazil) and FAPESC (Brazil) grants 2716/2012, TR 2012000344, and FAPESP (Brazil) grant 2013/24468-1. A.D. acknowledges the support from CNPq under grant 305639/2010-2. E.M. is supported by the Juan de la Cierva Program.}
}
%
%

\author{Eugenio Meg\'{\i}as\inst{1}\fnsep\thanks{\email{emegias@ifae.es}} \and
        D\'ebora P. Menezes\inst{2}  \and
        Airton Deppman\inst{3}
}

\institute{Grup de F\'{\i}sica Te\`orica and IFAE, Departament de F\'{\i}sica, Universitat Aut\`onoma de Barcelona, Bellaterra E-08193 Barcelona, Spain
\and
           Departamento de F\'isica, CFM, Universidade Federal de Santa 
Catarina, CP 476, CEP 88.040-900 Florian\'opolis - SC - Brazil
\and
           Instituto de F\'isica, Universidade de S\~ao Paulo - Rua do 
Mat\~ao Travessa R Nr.187 CEP 05508-090 Cidade Universit\'aria, S\~ao Paulo - 
Brazil 
          }

\abstract{%
We derive the nonextensive thermodynamics of an ideal quantum gas composed by bosons and/or fermions with finite chemical potentials. We find agreement with previous works when $\mu \le m$, and some inconsistencies are corrected for fermions when $\mu > m$. This formalism is then used to study  the thermodynamical properties of hadronic systems based on a Hadron Resonance Gas approach. We apply this result to study the protoneutron star stability under several conditions.
}
\maketitle
\section{Introduction}
\label{sec:intro}

Tsallis statistics constitutes a generalization of Boltzmann-Gibbs (BG) statistical mechanics, under the assumption that the entropy of the system is non additive. For two independent systems $A$ and $B$
\begin{equation}
S_{A+B} = S_A + S_B + (1-q)S_A S_B \,,
\end{equation}
where the entropic index $q$ measures the degree of nonextensivity~\cite{Tsallis:1987eu}. Many applications of this statistics have been done in the past few years. In high energy physics the distribution of transverse momentum is more naturally described with Tsallis statistics~\cite{Bediaga:1999hv}, and a non extensive generalization of the Hagedorn's theory pointed out that the hadronic systems in the confined phase are characterized by the value $q \approx 1.14$~\cite{Marques:2012px}. In this work we extend this statistics to finite chemical potential systems, and consider as a natural application the study of protoneutron stars. Until recently most equations of state (EoS) were expected to produce maximum stellar masses of the order of $1.4\, M_\odot$ and radii $13\,\Km$. However, the recent confirmation of two stars with masses $2\, M_\odot$ imposes more rigid constraints on the EoS~\cite{Demorest:2010bx}. In this communication we will explore which modifications the Tsallis statistics can introduce in the EoS of hadronic matter and protoneutron stars.

\section{Nonextensive thermodynamics at finite chemical potential}
\label{sec:nonextensive}

Tsallis' maximum entropy distributions can be conveniently written in terms of $q$-deformed functions, which in the limit $q\to 1$ reduce to the standard results in BG statistics. If we define the $q$-exp function $e_q^{(\pm)}(x)=[1 \pm (q-1)x]^{\pm1/(q-1)}\,$ for $x\ge 0$ $(x < 0)$, and the $q$-log function $\log^{(\pm)}_q(x)=\pm (x^{\pm(q-1)}-1)/(q-1)$, then the $q$-deformed quantum distribution functions have the form~\cite{Conroy:2010wt,Cleymans:2011in}
\begin{equation}
n_q^{(\pm)}(x) = \frac{1}{(e_q^{(\pm)}(x)-\zeta)^{\tilde{q}}} \,, \qquad \textrm{with} \qquad \tilde{q} =  
\begin{cases}
& q \quad\;\;\;\, \,,\quad\,\,\,x\geq 0 \\
& 2-q \; \,,\quad\,\,\,x<0
\end{cases}\,,
\end{equation}
where~$x = (E_p - \mu)/T$, and the particle energy is $E_p = \sqrt{\vec{p}^2+m^2}$ with $m$ the mass and $\mu$ the chemical potential.  In the following we take $\zeta = \pm 1$ for bosons and fermions respectively. The grand-canonical partition function for a nonextensive ideal quantum gas is defined as~\cite{Megias:2013upa}
\begin{equation}
 \log\Xi_q(V,T,\mu) =
 -\zeta V\int \frac{d^3p}{{(2\pi)^3}} \sum_{r=\pm}\Theta(r x)\log^{(-r)}_q\bigg(\frac{ e_q^{(r)}(x)-\zeta}{ e_q^{(r)}(x)}\bigg) \,, \label{partitionfunc}
\end{equation}
where $\Theta$ is the step function.  Eq.~(\ref{partitionfunc}) reduces to the Bose-Einstein and Fermi-Dirac partition functions in the limit $q \rightarrow 1$. The two sectors in the integrand of Eq.~(\ref{partitionfunc}), $x \ge 0$ and $x<0$, translate in momentum variable into $p \ge (\mu^2-m^2)^{1/2}$ and $p <  (\mu^2-m^2)^{1/2}$ respectively. The partition function for bosons is defined only for $\mu \le m$, therefore the sector $x<0$ contributes only for fermions when~$\mu > m$.

The thermodynamic functions can be obtained from Eq.~(\ref{partitionfunc}) by using the standard thermodynamic relations. At this point one must take care of the fact that the integrand in Eq.~(\ref{partitionfunc}) is a discontinuous function in $x=0$, i.e. in $p = (\mu^2-m^2)^{1/2}$. The location of the discontinuity is a function of the chemical potential, and as a consequence the derivatives with respect to $\mu$ induce contributions proportional to the jump of the integrand at the discontinuity. In particular, the average number of particles is
\begin{equation}
     \langle N \rangle = T\frac{\partial}{\partial \mu} \log \Xi_q \bigg|_T = V \left[ C_{N,q}(\mu,T,m) +  \int \frac{d^3p}{(2\pi)^3} \sum_{r=\pm} \Theta(r x) \bigg(\frac{1}{e_q^{(r)}(x) -\xi }\bigg)^{\tilde{q}}  \right]\,, \label{occnumb}
\end{equation}
where the momentum-independent term~$C_{N,q}(\mu,T,m) = \frac{1}{2\pi^2}T\mu(\mu^2-m^2)^{1/2} \frac{2^{q-1} + 2^{1-q} -2}{q-1} \Theta(\mu-m)$ is the contribution induced by the discontinuity~\cite{Megias:2013upa}. A similar term appears also in the average energy~$\langle E\rangle$. The entropy can be obtained through the relation $S = \frac{\partial}{\partial T}\left(T\log \Xi_q\right) \Big|_\mu$, resulting
\begin{equation}
S = V\!\int \!\!\frac{d^3p}{(2\pi)^3} \sum_{r=\pm} \!\!\Theta(r x) \bigg[ \!-\!n_q^{(r)}(x) \log_q^{(-r)}\left(\bar{n}_q^{(r)}(x)\right) + \zeta [1 + \zeta \bar{n}_q^{(r)}(x) ]^{\tilde{q}} \log_q^{(-r)}\left(\!1 + \zeta \bar{n}_q^{(r)}(x)\right) \!\!\bigg] \,, \label{eq:entropy}
\end{equation} 
where $\bar{n}_q^{(r)}(x) \equiv [n_q^{(r)}(x)]^{1/\tilde{q}}$. The integrands in Eqs.~(\ref{occnumb}) and (\ref{eq:entropy}) are also discontinuous functions in $x=0$. We have verified the thermodynamic consistency of these expressions by checking that  $\left( \partial S/\partial E\right)_{V,N} = 1/T$. In particular, if $C_{N,q}$ were dropped off, the thermodynamic consistency wouldn't be preserved for $\mu>m$. This result for the entropy agrees with~\cite{Conroy:2010wt} either for $x \ge 0$ or $x<0$, but it disagrees with~\cite{Cleymans:2011in} for $x<0$. The latter reference is however focused on applications to ultra relativistic collisions for which $T, m \gg \mu$, and so the sector $x<0$ doesn't play any role in this case. Nevertheless this sector is very important in the study of protoneutron stars. In addition, the inclusion of $C_{N,q}$ corrects some thermodynamical inconsistencies in the literature related to $\langle N \rangle$ and $\langle E \rangle$.

\section{Thermodynamical properties of hadronic systems}
\label{sec:thermo}

\begin{figure*}
\centering
\includegraphics[width=6.5cm,clip]{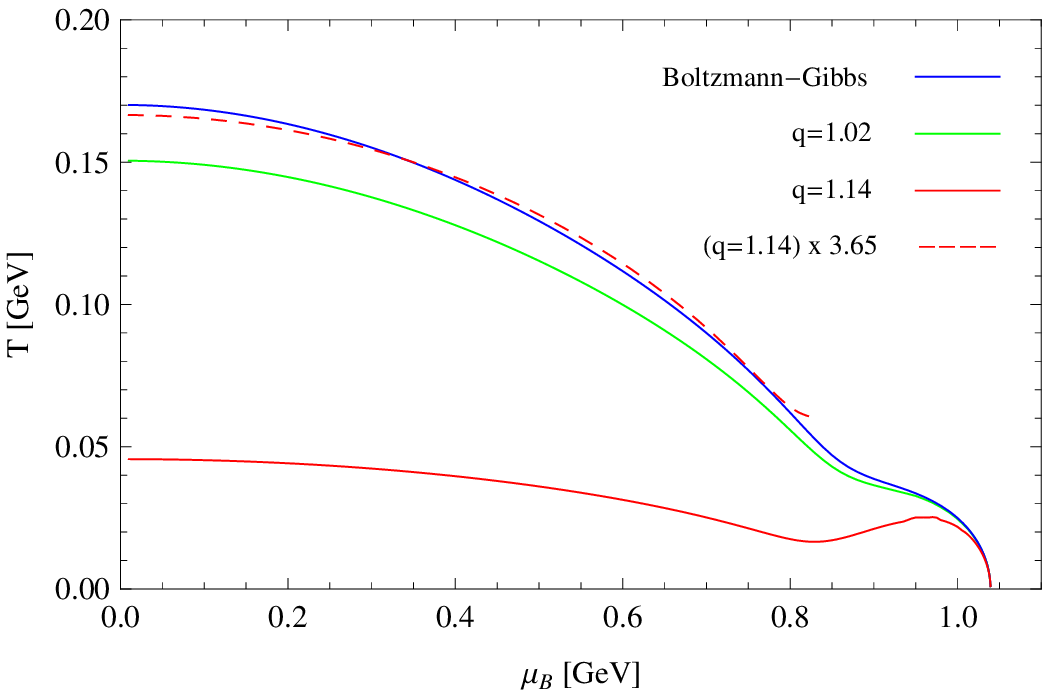}
\hspace{0.5cm}
\includegraphics[width=6.5cm,clip]{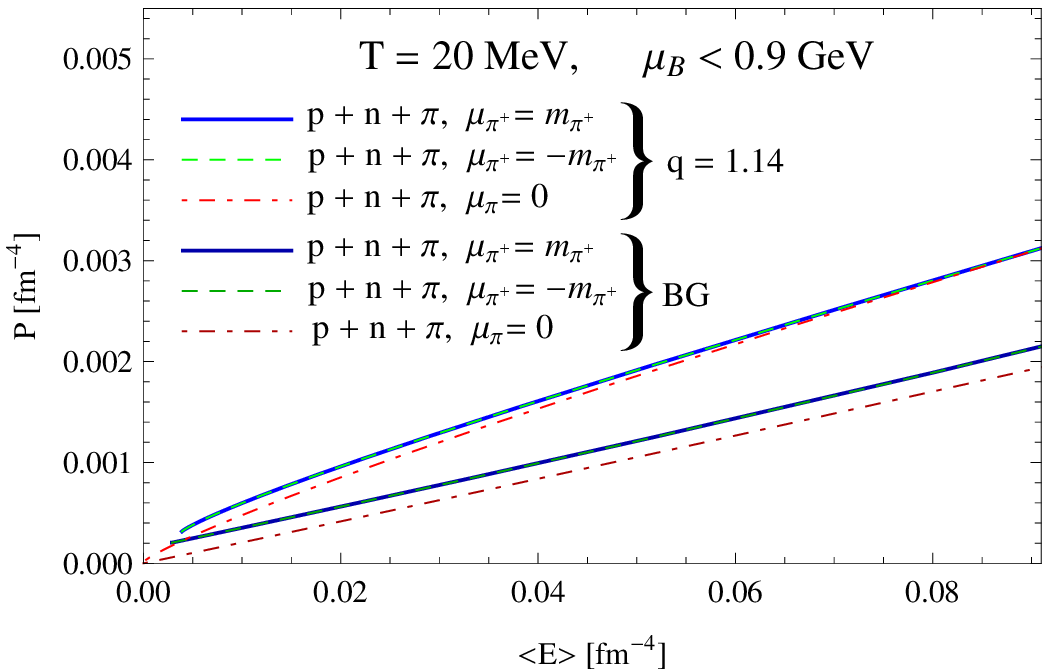}
\vspace{-0.1cm}
\caption{Chemical freeze-out line~$T = T(\mu_B)$ (left) and EoS of hadronic matter (right).}
\label{fig:Chemical_freeze_out}       
\end{figure*}

A useful and fruitful approach to study the thermodynamics in the confined phase of Quantum Chromodynamics (QCD) is the Hadron Resonance Gas model. This is based on the assumption that physical observables in this phase admit a representation in terms of hadronic states which are treated as non-interacting particles~\cite{Hagedorn:1984hz,Karsch:2003vd,Megias:2012kb,Megias:2013xaa}. These states are taken as the conventional hadrons listed in the review by the Particle Data Group~\cite{Beringer:1900zz}. Within this approach the partition function is given by
\begin{equation}
\log \Xi_q(V,T,\{\mu\})=\sum_i \log \Xi_q(V,T,\mu_i)\,, \label{eq:logZ}
\end{equation}
where $\mu_i$ is the chemical potential for the {\it i-th} hadron. Using the arguments of~\cite{Cleymans:1999st}, the phase transition line between confined and deconfined regimes of QCD in the $T \times \mu_B$ diagram can be determined by the condition $\langle E \rangle/\langle N \rangle=1 \, \GeV$. The total numbers of hadronic states we consider are 808 for mesons and 1168 for baryons ($+$ anti-baryons), corresponding to masses below $11\,\GeV$ and $5.8\,\GeV$ respectively, and we restrict the summation in Eq.~(\ref{eq:logZ}) to zero strangeness. The result is displayed in Fig.~\ref{fig:Chemical_freeze_out} (left).
In Tsallis statistics the temperature $\tau$ is an effective parameter
linearly related to the physical temperature $T$~\cite{Deppman:2012qt}. For $\mu_B=0$ the effective temperature is $\tau_o=45.6\,\MeV$ when $q=1.14$, which is in agreement within $25\%$ with the value obtained from the analysis of the $p_T$-distributions in high energy $p+p$ collisions~\cite{Marques:2012px}. The chemical freeze-out lines spam over the region $0<\mu_B<1039.2\,\MeV$.
\begin{figure*}
\centering
\includegraphics[width=7cm,clip]{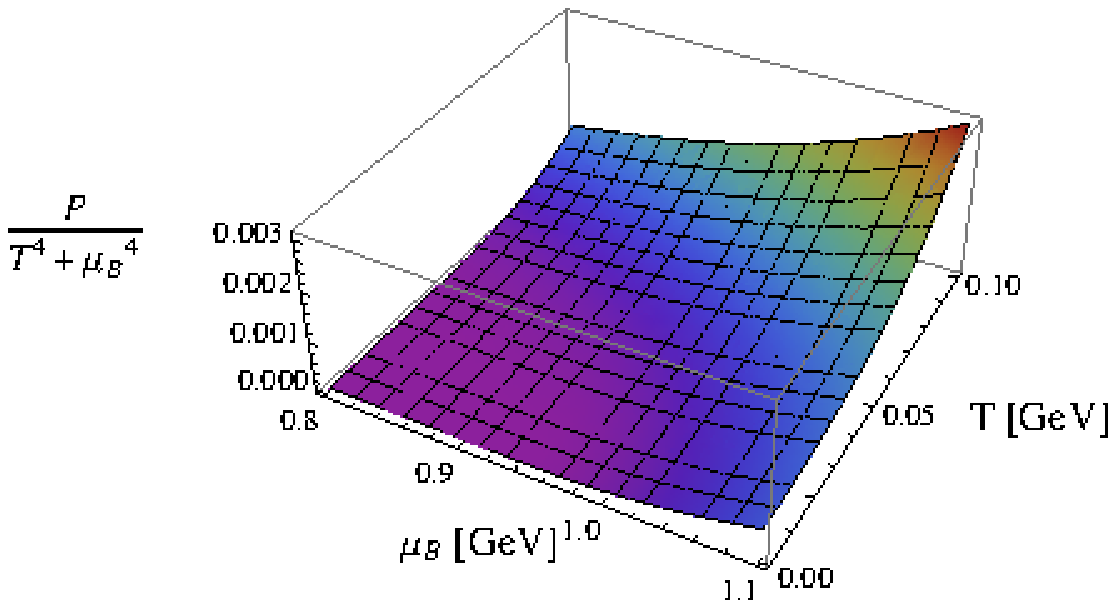}
\includegraphics[width=7cm,clip]{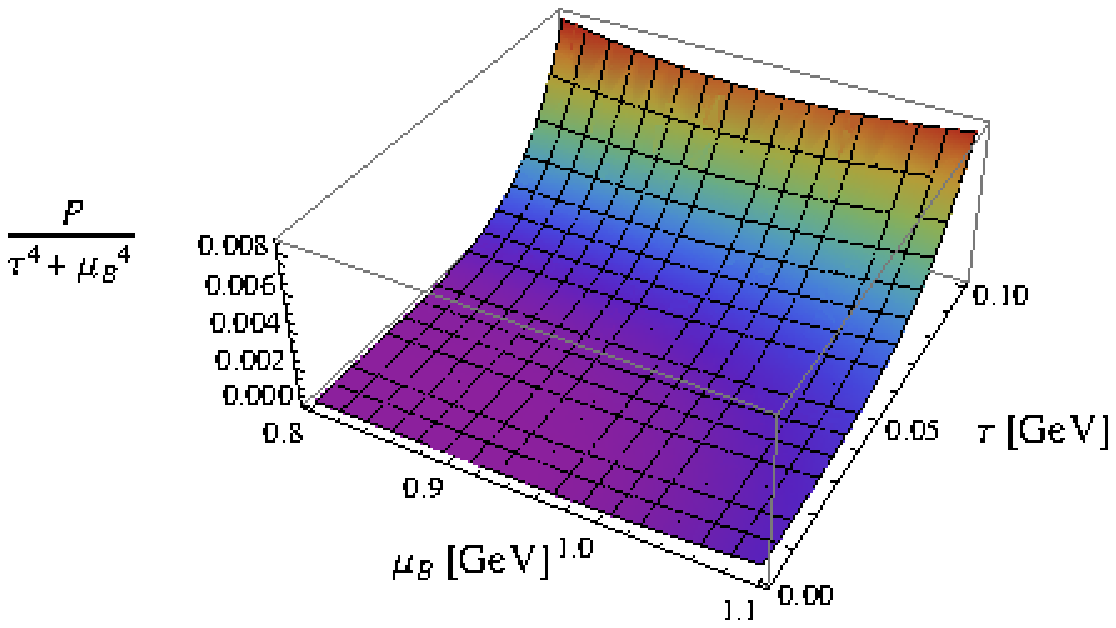}
\vspace{-0.1cm}
\caption{Pressure vs temperature and baryonic chemical potential. BG (left), Tsallis $q = 1.14$ (right).}
\label{fig:Pressure3D}       
\end{figure*}

We plot in Fig.~\ref{fig:Chemical_freeze_out} (right) the EoS for hadronic matter at finite baryonic chemical potential. It is remarkable that it becomes harder ($P(E)$ is larger) in Tsallis statistics as compared to BG statistics~\cite{Megias:2013upa}. The effect of considering a nonzero pion chemical potential is to produce an even harder EoS in both statistics. Fig.~\ref{fig:Pressure3D} shows a more rapid growth of the pressure in Tsallis statistics as compared to BG for all values of $\mu_B$, and this effect is much stronger for $\mu_B \sim 0.8\,\GeV$.

\section{Application to (proto)neutron stars}
\label{sec:stars}

In addition to hadronic matter, leptons are also present in stellar matter and they play an important role in the equilibration between gravitational force and the gas pressure due to their small masses. Our goal is to study the effects of Tsallis statistics as compared to BG statistics, and for this a constant temperature scenario $\tau \approx 20 \,\MeV$ is good enough.\footnote{We will study a fixed entropy scenario in a forthcoming work~\cite{Menezes:2014}.} We consider that the star is composed by neutrons, protons, pions, electrons and muons. The star is a dynamically equilibrated system because protons and neutrons are being converted into one another either through weak decays or scattering with pions,
\begin{equation}
 \begin{cases}
  & n \rightarrow p + e^-  + \bar{\nu}_e \\
  & p \rightarrow n + e^+ + \nu_e
 \end{cases} \,,
\hspace{3cm}
 \begin{cases}
  & n + \pi^+\rightarrow p \\
  & p + \pi^- \rightarrow n  
 \end{cases} \,.
\end{equation}
The dynamical equilibrium between the relative number of particles leads to the following relations between chemical potentials: $\mu_n = \mu_p + \mu_e$, $\mu_n = \mu_p - \mu_{\pi^+}$ and $\mu_e=\mu_{\mu}$. As the star is electrically neutral, the particles density condition $\rho_p + \rho_{\pi^+} = \rho_{\pi^-} + \rho_e + \rho_\mu$ is also enforced. Tsallis statistics has only been checked in hadron physics, and this justifies that we use it for hadrons only while the thermodynamics of leptons is described with the BG statistics for a fermionic free gas.

\begin{figure*}
\centering
\includegraphics[width=7cm,clip]{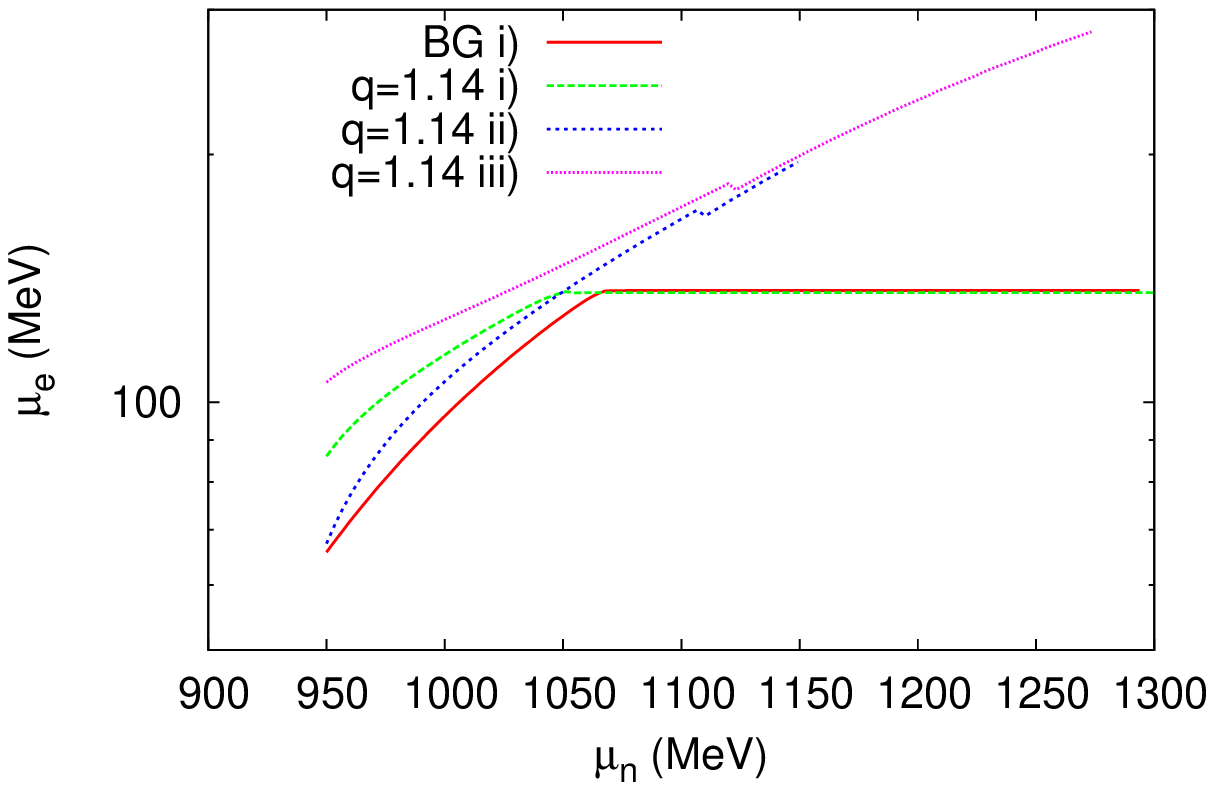}
\includegraphics[width=7cm,clip]{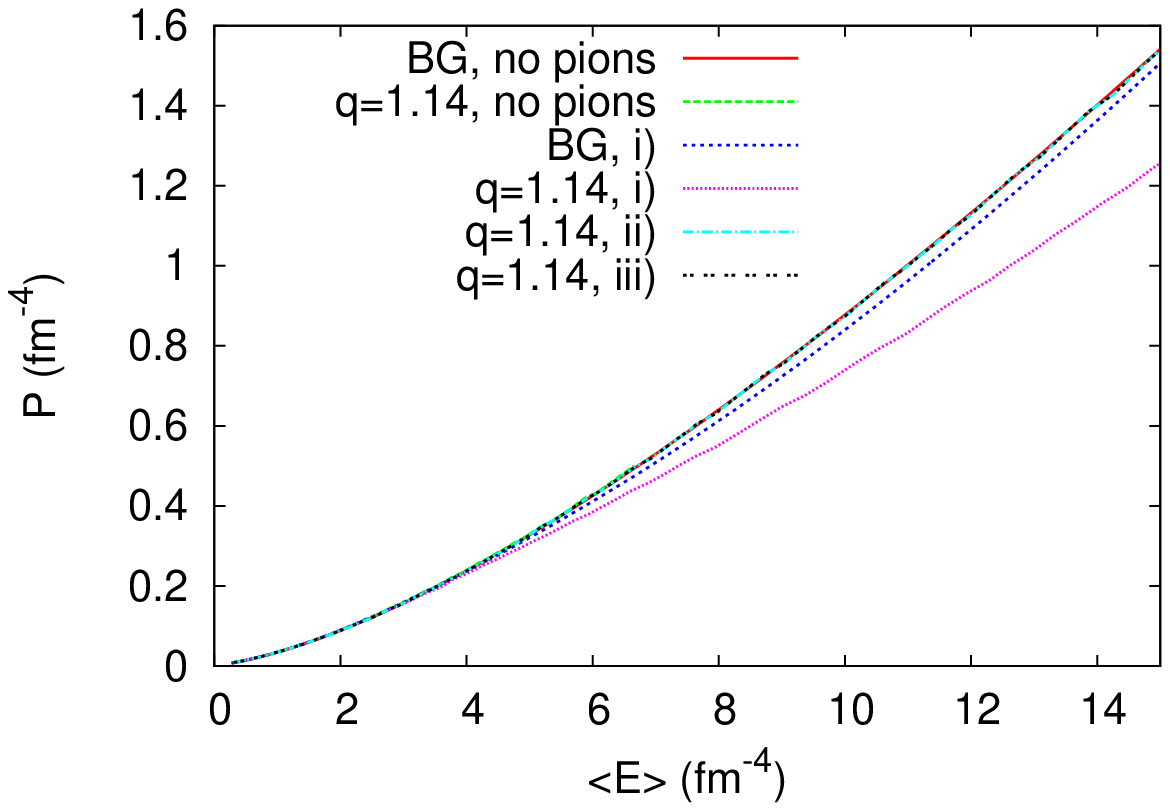}
\vspace{-0.3cm}
\caption{ $\mu_e$ vs $\mu_n$ (left) and EoS (right) for: i) $\mu_{\pi^-}=\mu_e$, ii) $\mu_{\pi^-}=m_{\pi}$ and  iii) $\mu_{\pi^-}=-m_{\pi}$.}
\label{fig:chemicals}       
\end{figure*}

\begin{figure*}
\centering
\includegraphics[width=7cm,clip]{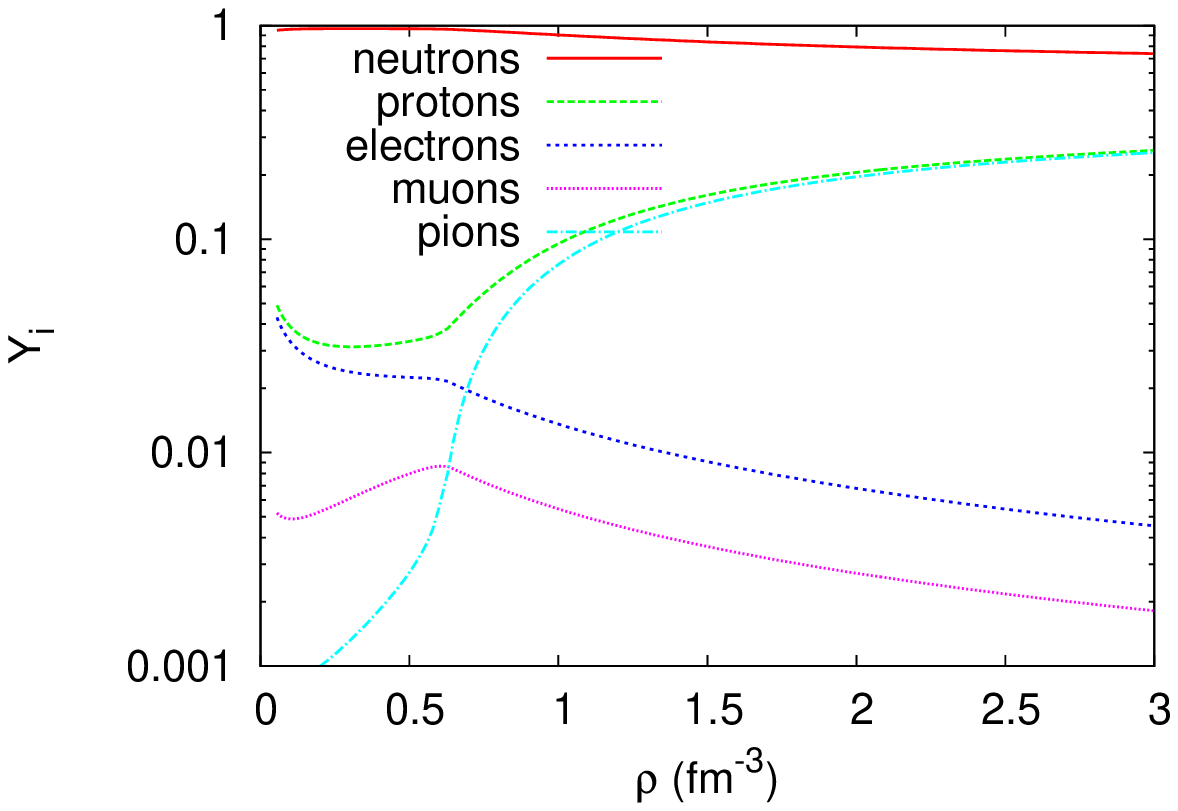}
\includegraphics[width=7cm,clip]{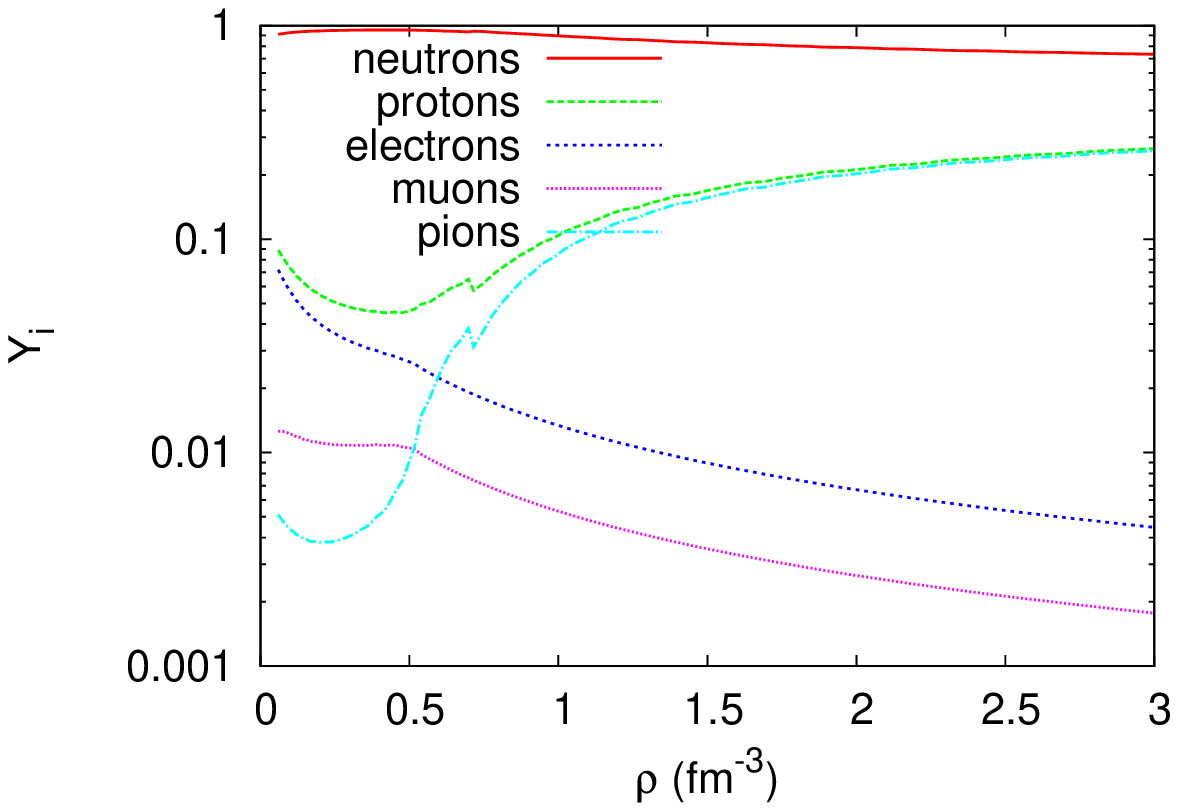}
\vspace{-0.3cm}
\caption{Particle fractions $Y_i \equiv \rho_i/\rho$ at $\tau = 20\, \MeV$ for $\mu_{\pi^-}=\mu_e$. BG (left), Tsallis $q=1.14$ (right).}
\label{fig:fractions}       
\end{figure*}

The result for chemical potentials, EoS and particle fractions are displayed in Figs.~\ref{fig:chemicals} and \ref{fig:fractions}.
Tsallis statistics leads to a discontinuity in the chemical potentials which is related to the discontinuity of the particle density integrals, Eq.~(\ref{occnumb}), when the momentum reaches the values $(\mu_{p,n}^2-m_{p,n}^2)^{1/2}$. The pion contribution can vary considerably, depending on the way its chemical potential is fixed, and in general Tsallis statistics increases the appearance of pions at low densities. While the EoS of hadronic matter obtained with $\mu_{\pi^+} = m_\pi$ is the hardest of all as it can be seen in Fig.~\ref{fig:Chemical_freeze_out} (right), the enforcement of charge neutrality and $\beta$-equilibrium wash out this effect, cf. Fig.~\ref{fig:chemicals} (right). By solving the Tolman-Oppenheimer-Volkoff equations~\cite{Oppenheimer:1939ne} with these EoS, the maximum stellar masses and radii lie around $0.7 \,M_\odot$ and $8 \Km$. This agrees with most of previous models~\cite{Menezes:2003pa,Lavagno:2011zm} that do not take nuclear interaction into account and, for this reason, it is not enough to explain the maximum values observed in nature~\cite{Demorest:2010bx}. A study taking into account these and other effects will be presented in~\cite{Menezes:2014}.

\section{Conclusions}
\label{sec:conclusions}

We have studied nonextensive thermodynamics including finite chemical potentials, and obtained fully thermodynamically consistent expressions for a quantum gas of bosons and fermions. Some inconsistencies in previous works have been addressed for~$\mu > m$ and for fermions. This formalism has been applied to study the thermodynamics of QCD, and we conclude that the EoS of hadronic matter becomes harder in Tsallis statistics as compared to BG statistics. The application to (proto)neutron stars shows that the presence of leptons washes out this hardening effect, and Tsallis statistics only cannot explain the star stability as observed in nature. Nevertheless nonextensive effects modify substantially the particle constitution at low densities, and this deserves more investigation.

%
%
%

\end{document}